\documentclass[a4paper,12pt]{article}
\usepackage[utf8]{inputenc}
\usepackage{cancel}
\usepackage{ulem}
\usepackage{amsfonts}
\usepackage{amssymb}
\usepackage{graphicx}
\usepackage{amsmath}
\usepackage{enumerate}
\usepackage{mathtools}
\usepackage{subfig}
\usepackage{color}
\usepackage{tikz}\usetikzlibrary{calc}
\usepackage{float}
\usepackage{here}
\usepackage{cite}
\usepackage{mathrsfs}
\usepackage{float,epsfig}
\usepackage{dcolumn}

\usepackage{graphicx}
\usepackage{bm}
\usepackage{amsmath,amssymb,amsthm}
\usepackage[colorlinks=true,linkcolor=blue,citecolor=red]{hyperref}
\newcommand{\be}{\begin{equation}}
\newcommand{\ee}{\end{equation}}
\newcommand{\bea}{\setlength\arraycolsep{2pt} \begin{eqnarray}}
\newcommand{\eea}{\end{eqnarray}}

\def\0{{\sst{(0)}}}
\def\1{{\sst{(1)}}}
\def\2{{\sst{(2)}}}
\def\3{{\sst{(3)}}}
\def\4{{\sst{(4)}}}
\def\5{{\sst{(5)}}}
\def\6{{\sst{(6)}}}
\def\7{{\sst{(7)}}}
\def\8{{\sst{(8)}}}
\def\sst#1{{\scriptscriptstyle #1}}

\makeatletter \@addtoreset{equation}{section}

\usepackage{multirow}
\definecolor{lime}{HTML}{A6CE39}

\begin{document}
%

\title{\normalsize
{\bf \Large	 Deflection Light  Behaviors  by  AdS Black Holes}}
\author{   \small A. Belhaj
\footnote{a-belhaj@um5r.ac.ma},
H. Belmahi \thanks{hajar\textunderscore belmahi@um5.ac.ma},
 M. Benali
 \thanks{mohamed\textunderscore benali4@um5.ac.ma}
\footnote{Authors in alphabetical order.}
	\hspace*{-8pt} \\
	{\small  D\'{e}partement de Physique, \'Equipe des Sciences de la mati\`ere et du rayonnement,
		ESMaR}\\ {\small   Facult\'e des Sciences, Universit\'e Mohammed V de Rabat}\\{\small Rabat, Morocco} \\
	{}
 }

 \maketitle
\begin{abstract}
	We investigate the behavior of the deflection of  light rays  by    charged and rotating AdS black holes   using the Gauss-Bonnet formalism.   Taking  weak field approximations and certain  appropriate limits associated with AdS geometries,    we compute and analyze   such an optical  quantity   by varying   the  involved moduli space parameters.  First, we study the charge  and the AdS radius effects on the deflection angle of  RN-AdS black holes.  For small values of the impact parameter $b$, we find that
the charge effect is  relevant.    Precisely, it decreases the deflection angle,  while the  AdS  background   one   is not.
For large values of $b$, however, these optical behaviors  have been  inverted   and the deflection angle becomes an increasing function of the charge. In this way, the cosmological
constant effect  is  remarked to be  relevant showing linear variations of the deflection angle.  Varying the charge,  we find  a  critical impact parameter  value $b_{c}$ where the charge effect is inverted.  For  rotating  solutions,  we show that the spinning parameter still decreases the deflection angle without any changing  behavior observed in  the charge effect.  Evincing of the cosmological constant, we recover  known results corresponding to charged and rotating  ordinary  black hole  solutions.   Examining  the plasma   effect, we reveal  that     the deflection angle   keeps the same behavior  being    a decreasing function in terms of  the frequency ratio.
		{\noindent}
		
	\end{abstract}
\newpage

\tableofcontents

\newpage

\section{Introduction}
The investigation of black hole physics has received a relevant interest in connections with various gravity models in different dimensions \cite{emparan,SWH,sss}. The associated studies have been encouraged and supported by  recent astrophysical observations \cite{A1,A2,A3}. Besides gravitational wave detections, this involves the imaging of a supermassive black hole candidate in the center of galaxy M87 provided by the Event Horizon Telescope (EHT) collaborations\cite{A4}. Concretely,    black hole physical aspects  have  been considered as  a rich subject  being approached using various methods and roads based  either on analytical or numerical computations. In particular, thermodynamic properties of AdS black holes have been extensively investigated by interpreting the cosmological constant as a pressure \cite{E1,F,F1}.  In this way, many transitions have been dealt with providing promising results \cite{12,f2}. A special emphasis has been put on the Hawking-Page transition of a large class of AdS black holes in arbitrary dimensions in the presence of nontrivial backgrounds including dark energy and D-brane objects embedded in higher dimensional supergravity models\cite{14,7,D}.  This gives results corresponding to the second order phase transition between large black holes (LBH) and small black holes (SBH). A  close examination shows that such a transition, based on the Gibbs free energy computations, has provided certain universalities in AdS  black hole physics\cite{E}. Moreover, optical behaviors of black holes have been also studied giving arise to interesting findings associated with shadow and deflection angle variations. In particular, it has been revealed that non-rotating black holes exhibit perfect circular geometries. The size of such shapes depends on black hole parameters including the mass, the charge and the stringy brane number\cite{Carlo,G,H,I,J}. These circular geometries have been distorted  by implementing the rotating parameter  generating non-trivial configurations called D-shapes, being controlled by astronomical observables describing  the size and the shape geometric deformations\cite{RCDM}. Among others, Kerr solutions with certain approximations could bring results matching with EHT collaborations \cite{K}.  In parallelled studies, the deflection angle of light rays has been  also investigated. This optical quantity could   unveil certain data which may impose particle physics constraints   associated with  black hole detections.  In particular, many works on such optical aspects  have been conducted\cite{4,M,DA1,5,1}. An inspection  shows that this optical quantity has been approached using different methods. Precisely, Gibbons and Werner have proposed a direct method to compute the deflection angle using the Gauss-Bonnet theorem. This method, which will be used in the present paper, is based on  backgrounds relying on a space metric \cite{L,DA2}. It is worth noting, in passing, that other methods have been exploited including the one based on the elliptic integral formalism combined with equations of motion \cite{N}. Concretely, it has been explored a link between the deflection angle and the thermodynamic behaviors of AdS black holes \cite{DATD}. Approaching  the deflection angle of black holes could bring new insights in the associated black hole physics.

The aim of this paper is to contribute to these activities by investigating  the deflection angle variations of certain charged and rotating AdS  black holes. This could support the recent findings which combine the thermodynamic and the deflection angle behaviors\cite{DATD}.  Using the Gauss-Bonnet formalism,  we compute and numerically analyze the  deflection angle   of AdS black hole solutions in terms of the involved physical parameters including the refractive index.  Analyzing the behavior of RN-AdS deflection angle, we show that for small values of the impact parameter $b$, 
the charge effect is  relevant. Concretely,  it decreases the deflection angle while the cosmological constant one is not.
For large values of $b$, however, these aspects have been  inverted. In this way, the cosmological
constant effect becomes relevant showing linear variations of the light  deflection angle. Moreover, the charge increases  the deflection angle.   For non-rotating solutions,  we observe  a specific  point  where the deflection angle behavior  gets changed   around  a critical  value of the impact parameter denoted by  $b_{c}$.  Moving to the rotating case,  we find that the rotating parameter decreases the deflection angle without any critical inverted point.  In the absence of  the cosmological constant, we recover    known results corresponding to charged and rotating  ordinary  black hole  solutions.   Inspecting the plasma   effect, we    find   that     the deflection angle  keeps the same behavior described by    a decreasing function in terms of the frequency ratio.

This work is structured as follows. In section 2, we give a concise review on deflection angle computations using the Gauss-Bonnet formalism. In section 3, we investigate   the deflection angle behaviors of charged AdS  black holes. Section 4 concerns the study of  charged and rotating AdS black holes. In section 4, we extend such results by considering the plasma medium effect. The last section  is devoted to  conclusions and      open questions.

 \section{Deflection angle of light  rays from  Gauss-Bonnet formalism}
 In this section, we present the formalism  being  needed  to examine  behaviors of the  deflection angle of  certain AdS black holes in four dimensions. A close inspection reveals that there are, a priori,  many ways to evaluate  such an optical quantity.  One of them  is based on the geodesic equations of motion controlling the dynamics of the studied   black holes. This method, however, brings   solutions in terms of complicated elliptic functions\cite{M,N}. The second method, which will be  exploited  here, relies on the Gauss-Bonnet theorem  and optic metric computations   \cite{DA6}.
 Considering the observer and the source  at finite distance in the equatorial plane, the deflection angle  can be expressed as
 \begin{equation}
 \Theta=\Psi_{R}-\Psi_{S}+\phi_{SR},
 \label{a1}
 \end{equation}
where $\Psi_{R}$ and $\Psi_{S}$ are angles between the light rays and the radial direction at the observer and  the source position,  respectively. $\phi_{SR}$ is the longitude separation angle \cite{DA6}. The computation of the deflection angle can be elaborated by exploiting the Gauss-Bonnet theorem \cite{L}. Indeed, the associated relation  is given by
\begin{equation}
\iint_{\Sigma} K dS + \sum_{\alpha=1}^{n}\int_{C_{\alpha}}k_{g}dl+\sum_{\alpha=1}^{n}\theta_{\alpha}=2\pi,
\label{a2}
\end{equation}
where $\Sigma$ is a two dimensional orientable  real surface bounded by curves $C_{\alpha} (\alpha=1,\ldots, n)$. $\theta_{\alpha}$ indicates the external angle at the $\alpha$-vertex. $K$  is  the Gaussian curvature of the surface $\Sigma$ and $k_{g}$ represents the geodesic curvature of $C_{\alpha}$. By considering the quadrilateral $^{\infty}_{r}\square^{\infty}_{S} $ formed of light curves   from the source  to the observer, it has been shown that Eq.(\ref{a1}) and Eq.(\ref{a2}) provide  a reduced integral formula on the space geometry  which reads
\begin{equation}
\Theta=-\iint_{^{\infty}_{r}\square^{\infty}_{S}} KdS+\int^{R}_{S}k_{g}dl.
\label{a3}
\end{equation}
It is noted that $dS$ is the area element of the involved surface.  The line element $dl$ can be handled by considering axisymmetric   black holes with the following   metric form
\begin{equation}
ds^2=-A(r,\theta)dt^2+B(r,\theta)dr^2+C(r,\theta)d\theta^2+D(r,\theta)d\phi^2-2H(r,\theta)dt d\phi.
\end{equation}
To defined the Riemannian manifold for which  the light geodesics are interpreted as spatial curves, we should exploit the null geodesic condition $ds^2=0$ \cite{DA6}. Indeed, this gives
\begin{equation}
dt=\pm\sqrt{\gamma_{ij}dx^{i}dx^{j}}+ \eta_{\phi}d\phi,
\end{equation}
where $\gamma_{ij}$ is a spatial metric. In the equatorial plane $(\theta=\pi/2)$ at constant $t$ of the space-time metric, one has a  2-dimensional curved space which is   represented by 
\begin{equation}
dl^2\equiv \gamma_{ij}dx^{i}dx^{j}.
\label{labe}
\end{equation}
In this  way,  the Gaussian curvature is expressed as follows
\begin{equation}
K=\frac{R_{r\phi r \phi}}{\gamma}=\frac{1}{\sqrt{\gamma}}\left( \frac{\partial}{\partial\phi}\left( \frac{\sqrt{\gamma}}{\gamma_{rr}}\Gamma^{\phi}_{rr}\right)  -  \frac{\partial}{\partial r}\left( \frac{\sqrt{\gamma}}{\gamma_{rr}}\Gamma^{\phi}_{r\phi}\right) \right) ,
\label{a7}
\end{equation}
where  one has used $\gamma=det(\gamma_{ij})$. The area element of the Eq.(\ref{a2}) takes the form 
\begin{equation}
dS=\sqrt{\gamma}drd\phi.
\end{equation} 
 According to \cite{DA6}, the geodesic curvature in this Riemannian manifold  is given by
\begin{equation}
k_{g}=-\frac{1}{\sqrt{\gamma \gamma^{\theta \theta}}}\eta_{\phi,r}.
\label{a28}
\end{equation}
Having presented the associated calculations, we move now to investigate certain optical aspects of AdS black holes by computing the corresponding deflection angle. Concretely, we approach such an optical quantity by varying the involved parameters in different backgrounds.
\section{Deflection angle   behaviors  by  Reissner–Nordström AdS black holes}
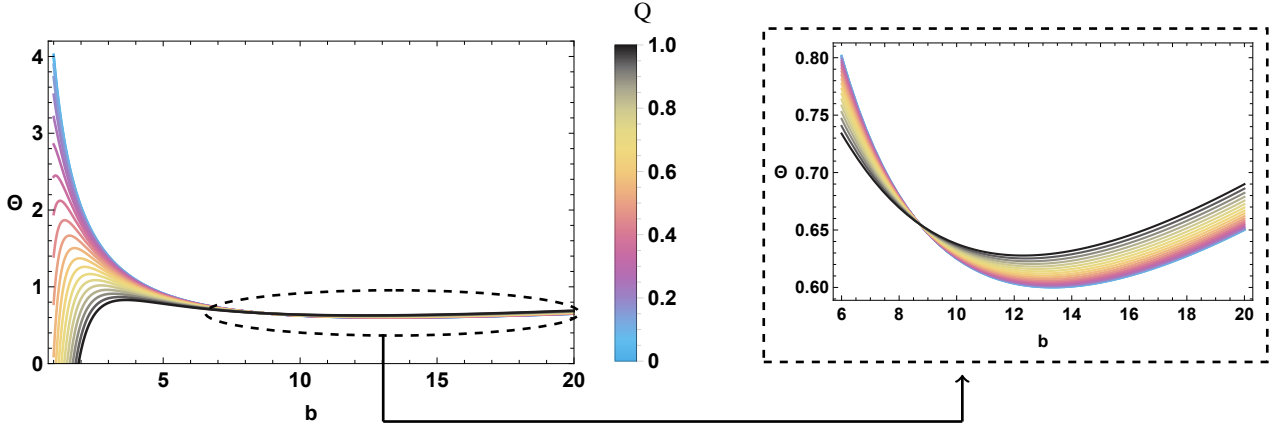
\begin{figure}[!ht]
		\begin{center}
		\begin{tikzpicture}[scale=0.2,text centered]
\node[] at (-20,1){\small  \includegraphics[scale=0.6]{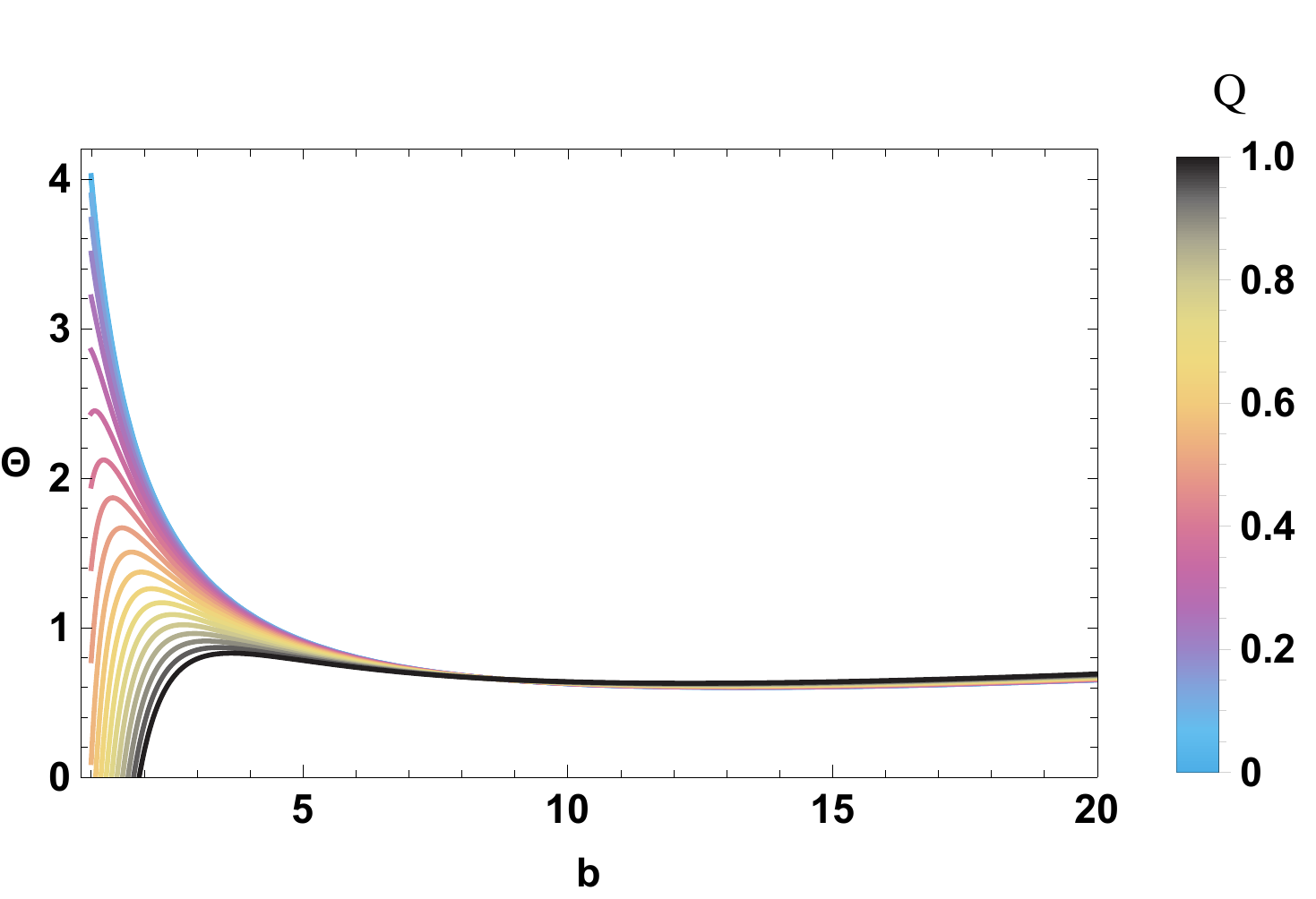}};	
\node[draw, line width=1pt,color=black,name=plan,dashed] at (24,1){\small  \includegraphics[scale=0.5]{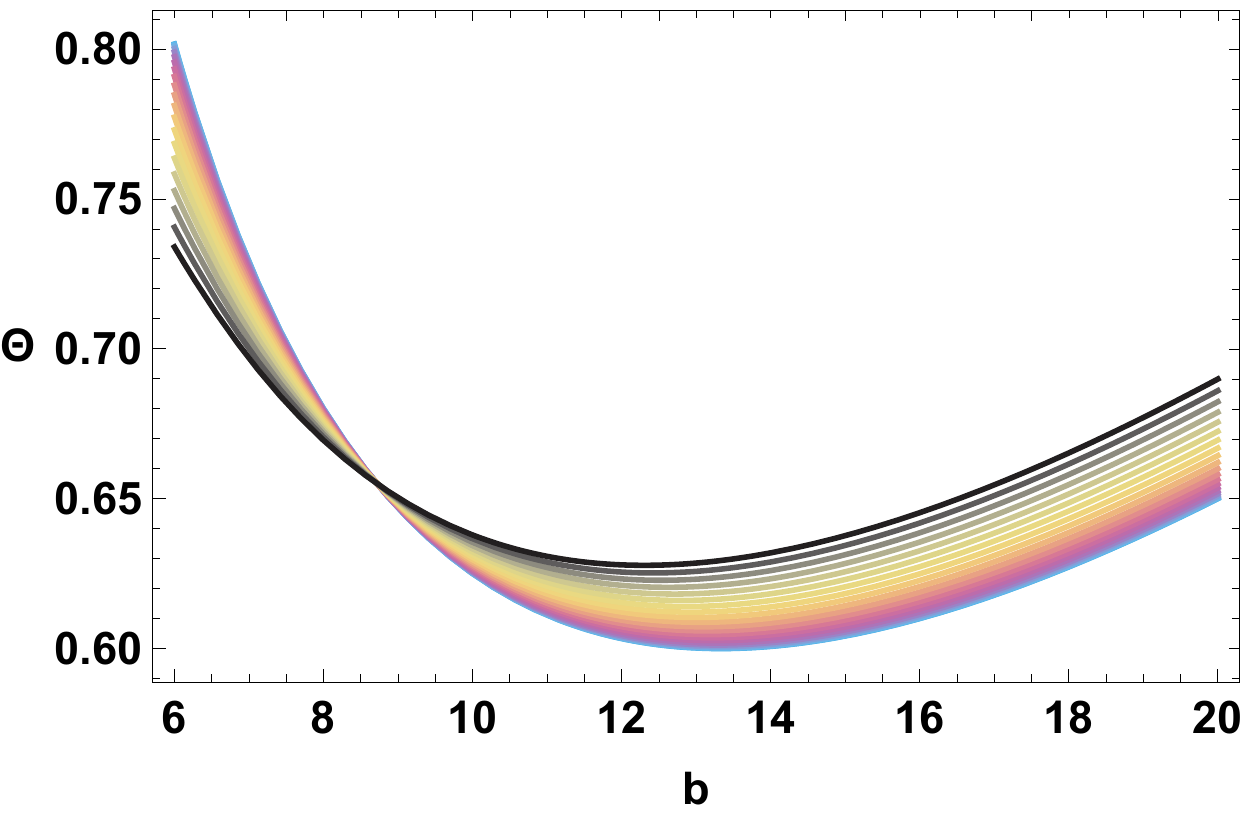}};	
\draw[-,line width=1pt,color=black](-17.6,-14)--(20.5,-14);
\draw[-,line width=1pt,color=black](-17.6,-14)--(-17.6,-8.2);
\draw[->,line width=1pt,color=black](20.5,-14)--(20.5,-10.9);
\draw[line width=1pt,dashed,rotate=90](-6.8,17) ellipse (1.5 and 12.3);
\end{tikzpicture}	
\caption{Variation of  the deflection angle  in terms of the impact parameter   for different  charge values by considering $u_{S}=u_{R}=0.1$, $\ell=20$ and  $M=1$. }
\label{im2}
\end{center}
\end{figure}

In this section, we are interested in  deflection angle behaviors of charged solutions called Reissner–Nordström  AdS (RN-AdS)   black holes in four dimensions. Indeed, the line element of the corresponding space-time metric is written as
\begin{equation}
ds^{2}=-f(r)dt^2+\frac{dr^2}{f(r)}+r^2(d\theta^2+\sin^{2}(\theta)d\phi^2),
\end{equation}
where $f(r)$ indicates the metric function given by $f(r)=1-\frac{2 M}{r}+\frac{r^2}{\ell^2}+\frac{Q^2}{r^2}$. $M$ and $Q$ represent the mass and the charge of AdS black holes, respectively. $\ell$ is the  AdS radius linked to four dimensional cosmological constant via the following relation
\begin{eqnarray}
\Lambda=-\frac{ 3}{\ell^2}.
\end{eqnarray}
 Applying the null geodesic conditions in the equatorial plane, one  gets the optical metric expression
\begin{eqnarray}
dt^2&=& \frac{1}{f(r)^2}dr^2+\frac{r^2}{f(r)}d\phi^2.
\label{a8}
\end{eqnarray}
 Using Eq.(\ref{a7}) and Eq.(\ref{a8}), the  Gaussian curvature  takes the form
\begin{equation}
 K\simeq\frac{1}{\ell^2}-\frac{6 M}{\ell^2 r}+\frac{6 Q^2}{\ell^2 r^2}-\frac{2 M}{r^3}+\frac{3 Q^2}{r^4}-\frac{6M Q^2}{r^5}+O(M^2,Q^3,1/\ell^{4}).
 \end{equation}
 In this case, the  determinant of the optical metric  can  be expressed as follows
 \begin{equation}
 \gamma={r^2}{\left(1-\left(\frac{2 M}{r} -\frac{r^2}{\ell^2}-\frac{Q^2}{r^2}\right) \right)^{-3}}
 \label{det}
 \end{equation}
which provides the following needed approximation
 \begin{equation}
 \sqrt{\gamma}\simeq r+O(M^1,Q^2,1/\ell^{2}).
  \label{detd}
 \end{equation}
 To keep the order $O(M^2,Q^3,1/\ell^{4})$  in deflection angle computations, we  use the  Eq.(\ref{detd}). Indeed, the computation gives
\begin{eqnarray} 
\int^{\phi_{R}}_{\phi_{S}}\int_{r_{o}}^{\infty} K \sqrt{\gamma}drd\phi&\simeq&\int^{\phi_{R}}_{\phi_{S}}\int_{r_{o}}^{\infty} drd\phi\;  (\frac{r}{\ell^2} +\frac{3 Q^2}{r^3}+\frac{6 Q^2}{\ell^2 r} -\frac{6 M}{\ell^2}-\frac{2 M}{r^2}-\frac{6 M Q^2}{r^4}\notag \\&+& O(M^2,Q^3,1/\ell^{4})
\label{a34}
 \end{eqnarray}
where $r_{0}$ is  the distance of the closest approach corresponding to the solution of the orbit equation. To develop such computations,   the impact parameter of motion is needed. It is given by
 \begin{equation}
 b=\frac{\mid L \mid}{E}=\frac{r^2}{f(r)}\frac{d\phi}{dt},
 \end{equation}
 where $E$ and $L$ are the two constants of motion. Putting $u=\frac{1}{r}$, the null geodesic condition gives
 \begin{equation}
 \left( \frac{du}{d\phi}\right) ^{2}=\frac{1}{b^2}-\frac{1}{\ell^2}-u^2 \left(-2 M u+Q^2 u^2+1\right).
 \label{a36}
 \end{equation}
To solve the second derivation of Eq.(\ref{a36}),  we should use the  perturbative method which provides that
 \begin{equation} u(\phi)=\frac{1}{b}\sin(\phi)+ O \left(M \right) .
 \label{a37}
 \end{equation}
In the weak field approximations and for small values of the AdS radius, the integral of   Eq.(\ref{a34}) can be expanded as follows
 \begin{eqnarray}
- \int^{\phi_{R}}_{\phi_{S}}\int_{r_{o}}^{\infty} K \sqrt{\gamma}drd\phi &\simeq& \int^{\phi_{R}}_{\phi_{S}}\int_{0}^{u(\phi)=\frac{1}{b}\sin(\phi)}2 M -3 Q^2 u+6 M Q^2 u^2 -\frac{1}{\ell^2 u^3}+\frac{6 M}{\ell^2 u^2}-\frac{6 Q^2}{\ell^2 u}du d\phi
 \notag \\
&\simeq & \frac{2M}{b}\left[ \sqrt{1-(bu_{S})^{2}}+\sqrt{1-(bu_{R})^{2}}\right]\notag \\&+&\left[ \frac{6 Q^{2}}{\ell^2}-\frac{3Q^{2}}{4b^{2}}\right] \left[\pi  - \arcsin(bu_{S})- \arcsin(bu_{R})\right]\notag \\&-&\frac{3Q^{2}}{4b^{2}}\left[bu_{R}\sqrt{1-(bu_{R})^{2}}+ bu_{S}\sqrt{1-(bu_{S})^{2}} \right] \notag\\&-&\frac{M Q^2}{3 b^3}\left[ \left( 16+(bu_{R})^{2}\right) \sqrt{1-(bu_{R})^{2}}+ \left( 16+(bu_{S})^{2}\right) \sqrt{1-(bu_{S})^{2}}\right] \notag \\&+& \frac{b}{2 \ell^2}\left[ \frac{\sqrt{1-(b u_{R})^2}}{u_{R}}+\frac{\sqrt{1-(b u_{S})^2}}{u_{S}}\right] -\frac{M b}{2\ell^2}\left[ \frac{1}{\sqrt{1-(b u_{S})^2}}+\frac{1}{\sqrt{1-(b u_{R})^2}}\right] \notag\\
&-&\frac{6 Q^{2}b}{\ell^2}\left[ u_{S}\arctan(u_{S}b)+u_{R}\arctan(u_{R}b))\right],
\label{a38}
 \end{eqnarray}
 where $u_{S}$ and $u_{R}$ are the inverse of the source and the observer distance. In this case, where the coefficient $H(r,\theta)$ of the  coupling term  $dtd\phi$
is missed, it has been observed  that the second term of the deflection angle expression associated with $k_g$ takes a vanishing value. The integral of  Eq.(\ref{a38})  can apparently diverge in the limit  $bu_{S}\to 0$ and   $bu_{R}\to 0$. This divergence  could be linked to the cosmological constant effect. Combining Eq.(\ref{a38}) and Eq.(\ref{a3}) and using the  asymptotic case  corresponding to  $bu_{S}<< 1$ and   $bu_{R}<< 1$, the deflection angle  of the RN-AdS black hole is found to be
\begin{equation}
\label{eq1}
\Theta\simeq\frac{4M}{b}-\frac{3 Q^2 \pi}{4 b^2}-\frac{32 M Q^2}{3 b^3}+\frac{6 Q^2 \pi}{\ell^2}-\frac{M b}{ \ell^2}+\frac{b}{2 \ell^2}\left[ \frac{1}{u_{R}}+\frac{1}{u_{S}}\right] +O(M^2,Q^3,1/\ell^{4}).
\end{equation}
An examination shows that one can make contact with known results.
Putting $\ell\rightarrow \infty$, we recover,  indeed, the expression of RN black holes \cite{4,DA5}. It has been observed that  the  expression of the deflection angle exhibits   a divergent behavior associated with vanishing limits of  $u_{S}$ and  $u_{R}$  functions. Taking $Q=0$, similar behaviors with the same terms of the AdS contributions have been found in \cite{DA4} for $r_{g}=2M$.
 \begin{figure}[!ht]
		\begin{center}
		\centering

			\includegraphics[width=9cm, height=8cm]{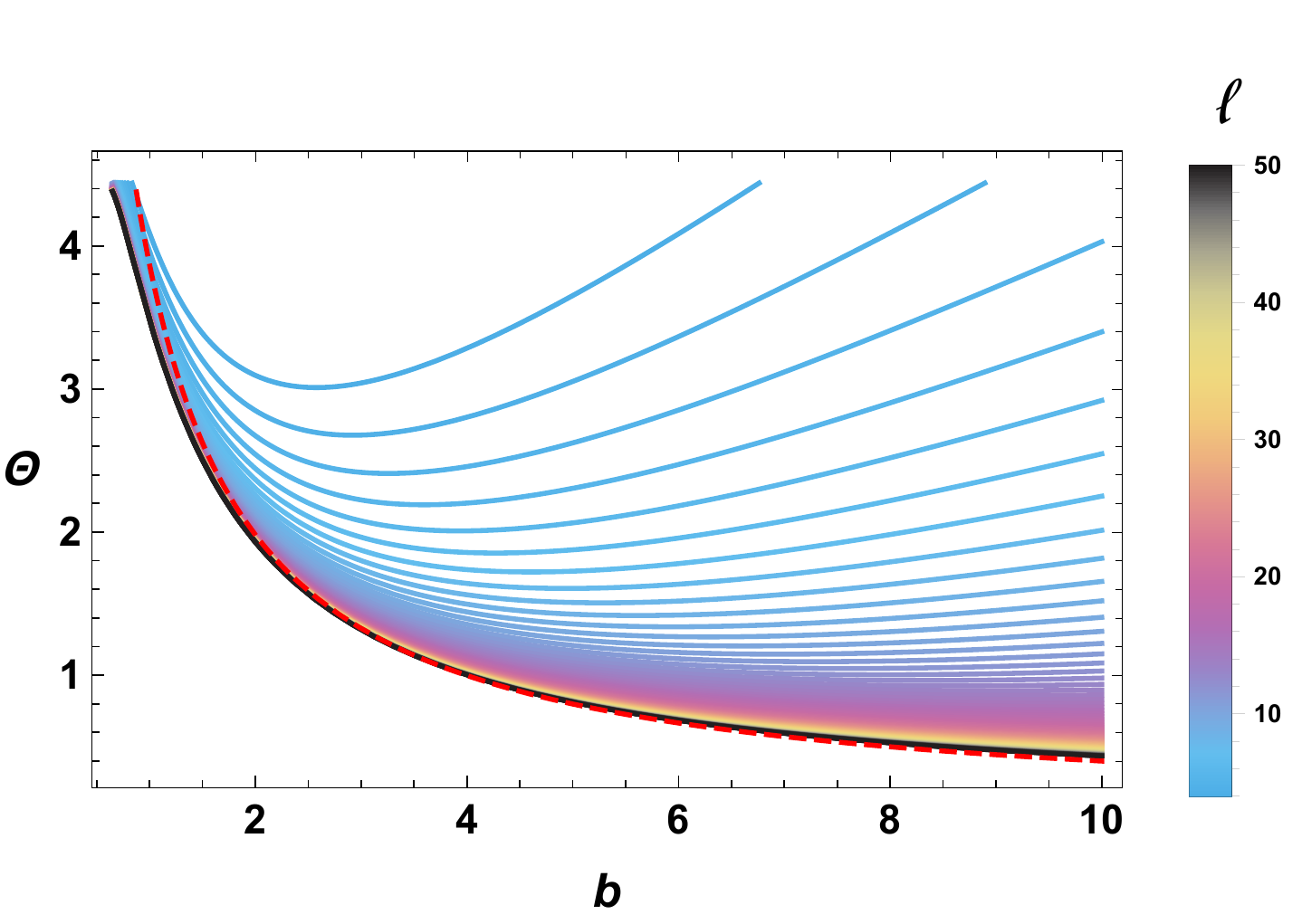}

\caption{Variation of the deflection angle in terms of the impact parameter for different values of $\ell$ by taking $u_{S}=u_{R}=0.1$, $M=1$ and $Q=0.2$. The dashed  red curve represents  the deflection angle of the  Schwarzschild black hole.}

\label{im1}
\end{center}
\end{figure}
 Firstly, we examine the behavior of the deflection angle as a function of the impact parameter $b$ in  AdS backgrounds. Considering the expression of  the deflection angle given by Eq.(\ref{eq1}), the variation of such an optical  quantity is illustrated in Fig.(\ref{im1}).  Fixing the mass and the charge,  we consider certain AdS radius values where theirs  contributions will be relevant. It follows from this figure that the deflection angle decreases rapidly for small values of the impact parameter. Then,  it increases by growing $b$. It is worth noting that similar behaviors have been  obtained in \cite{DA3}. Increasing the value of $\ell$, the deflection angle decreases.  It has been  revealed    that the presence of the AdS geometry  could   deviate the light rays. For small values of the impact parameter $b$, the deflection angle of the  Schwarzschild black hole  is a little large  compared with the one corresponding to AdS contributions. Fixing $\ell=50$  and taking  large values of $b$, we remark the  same behavior appearing in  the deflection angle of the Schwarzschild  black  hole solution. For  $\ell<50$, the deflection angle in the presence of the cosmological constant becomes relevant. In the Fig(\ref{im2}), we inspect the charge effect on the deflection angle of RN AdS black holes.  For small values of $b$,  the deflection angle deceases by increasing the charge. This behavior has been  inverted for large values of the impact parameter.  However, the inverted   behavior starts   from a specific value of  the impact parameter $b\simeq 8.716$  for $M=1$ and $\ell =20$. It is  follows from the figure that  all $Q$ curves meet at a critical point $b_{c}$.  This means that this critical value   should  depend only on $M$ and $\ell$.  A close  examination shows that an explicit expression of  $b_{c}$  can be obtained by using the derivation of the deflection angle  with respect to the charge. The computation gives
\begin{equation}
b_{c}=\frac{ (\lambda \sqrt{6})^{2/3}+(6 \pi )^{2/3} \ell ^2}{12( \pi\lambda)^{1/3} }
\end{equation}
where one has used $ \lambda=\left(\sqrt{65536 M^2 \ell ^4-6 \pi ^2 \ell ^6}+256 M \ell ^2\right)$.  It has been observed that this behavior changing of the deflection angle, in terms of  the charge,  appears  in the  case of the quintessential RN black hole\cite{5}. However, this is removed in the case of RN ordinary black holes. These behaviors could be   understood  from the metric function $f(r)$ expression. For small values of $b$, the charge effect is notable while the cosmological constant one is not relevant. In this case,  we observe that the deflection angle  involves a maximum. The letter is increasing by decreasing  the charge $Q$. For large values of $b$, these properties have been inverted. Indeed, in this case, the cosmological constant effect becomes relevant showing linear variations of the deflection angle of light rays. We expect to obtain similar optical aspects for a large class of AdS black holes in four dimensions.

  \section{Deflection angle of Kerr-Newman AdS black holes }
\begin{figure}[!ht]
		\begin{center}
		\begin{tikzpicture}[scale=0.2,text centered]
\node[] at (-45,1){\small  \includegraphics[scale=0.6]{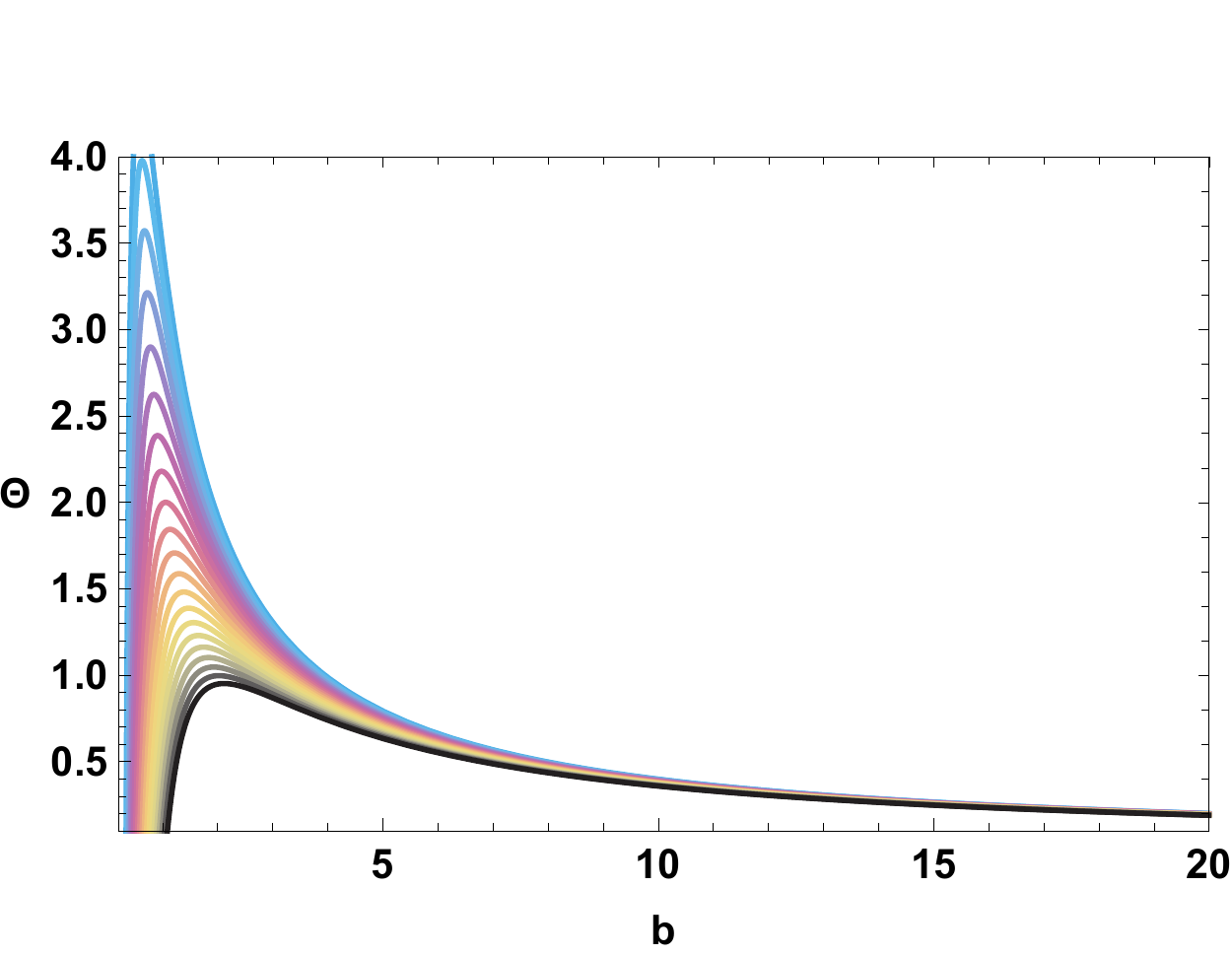}};	
\node[] at (0,1){\small  \includegraphics[scale=0.6]{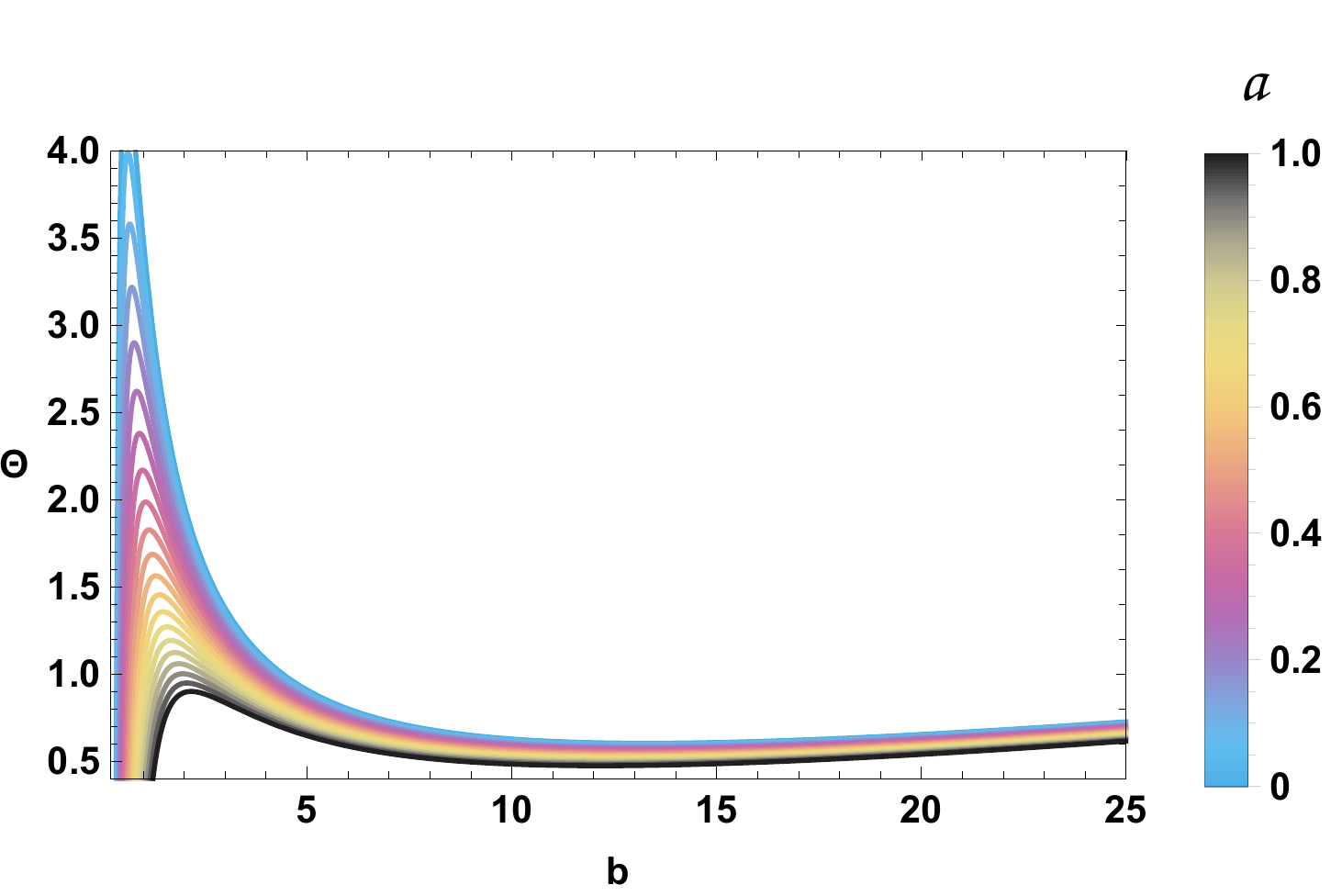}};	
\node[] at (3,3){\small  \includegraphics[scale=0.3]{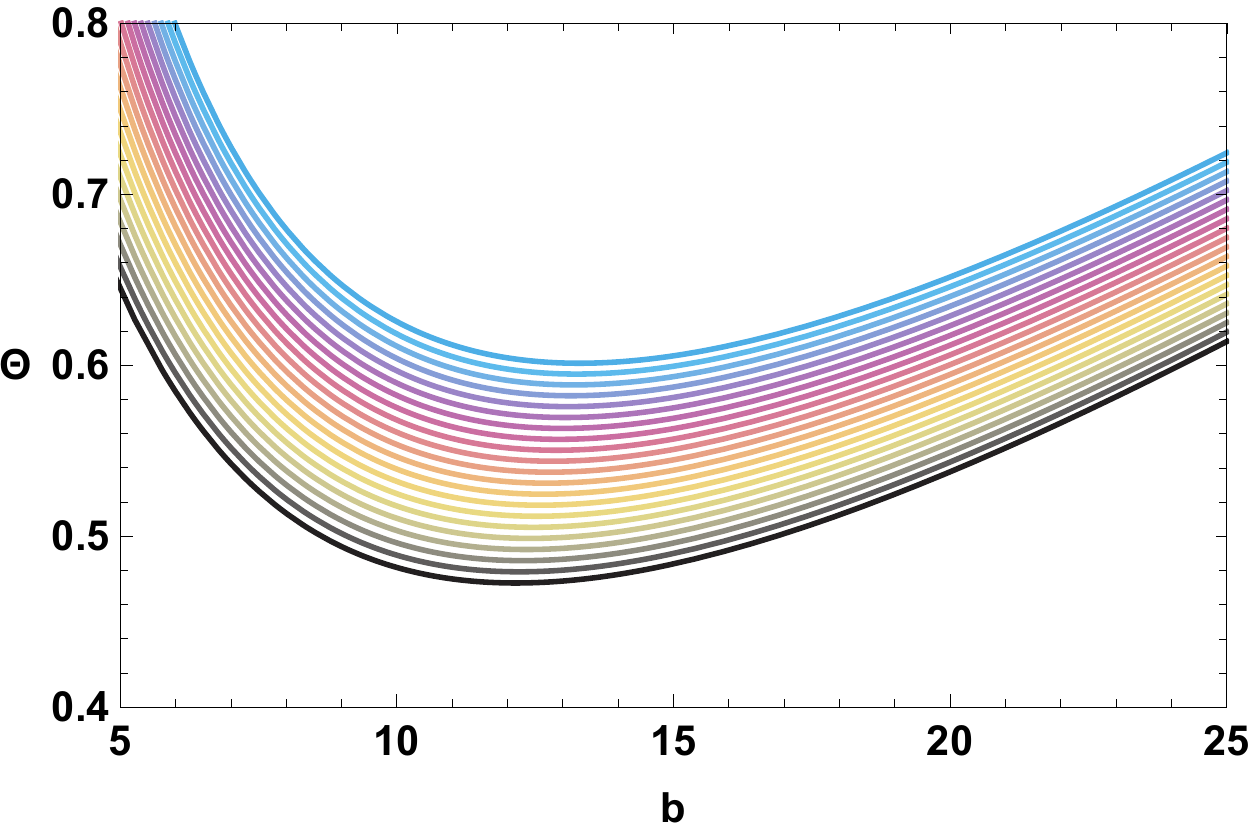}};	
\node[] at (-39,3){\small  \includegraphics[scale=0.3]{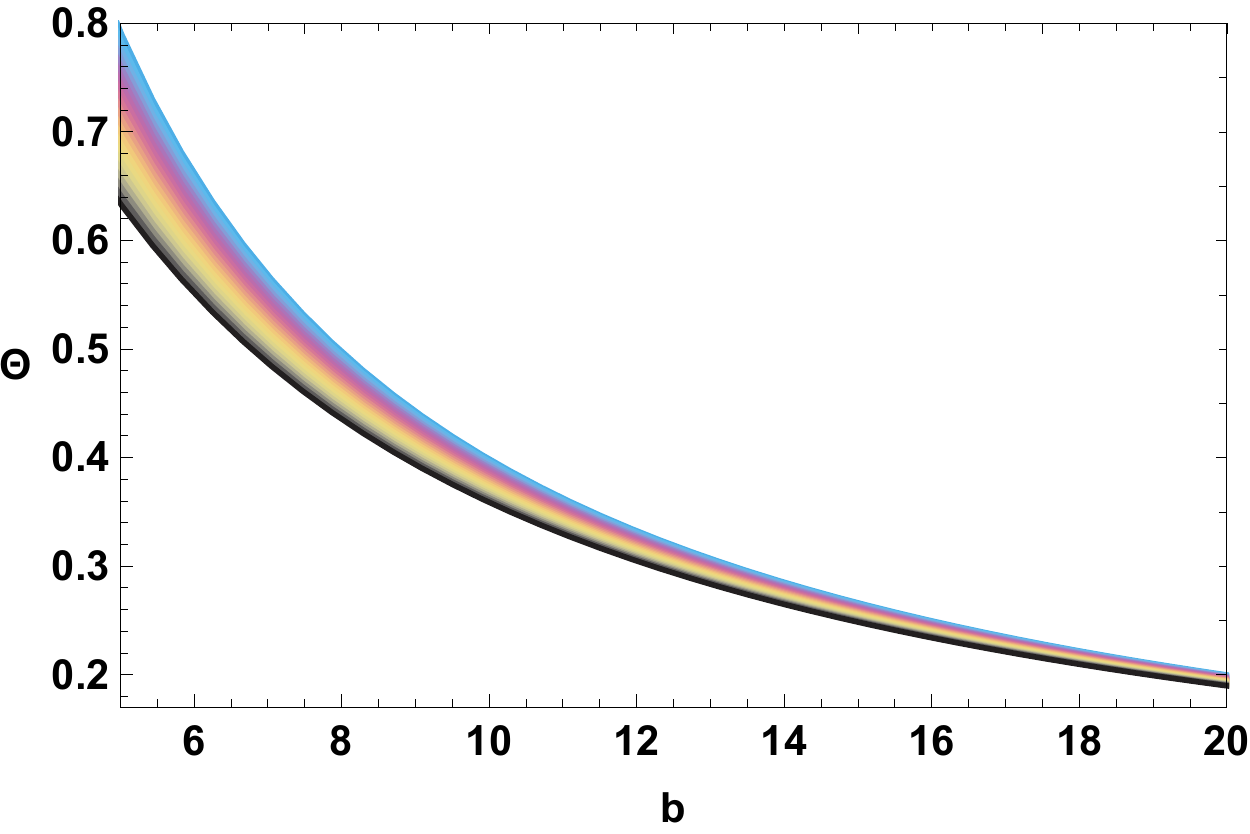}};	
\end{tikzpicture}	
\caption{Variation of the deflection angle in terms of the impact parameter for different values of $a$ by taking $u_{S}=u_{R}=0.1$ , M=1 and $Q=0.2$. Left panel: behavior with $\ell=\infty$, right panel: behavior with $\ell=20$. }
\label{im5}
\end{center}
\end{figure}

 Looking for the rotating parameter effect in the  AdS geometry, we investigate the deflection angle of  Kerr-Newman AdS black holes.  According to \cite{DAR1}, the line element of this class of black holes is given by
  \begin{eqnarray}
 ds^2&=&-\frac{\Delta_{r}}{\Sigma}\left[dt-\frac{a\sin^2\!\theta}{\Xi}d\phi\right]^2
+\frac{\Sigma}{\Delta} dr^2+\frac{\Sigma}{\Delta_{\theta}}d\theta^2+\frac{\Delta_{\theta}\sin^2\!\theta}{\Sigma}\left[a dt-\frac{r^2+a^2}{\Xi}d\phi\right]^2,
\end{eqnarray}
where one has
\begin{eqnarray}
\Sigma&=&r^2+a^2\cos^2\!\theta\,,\quad \Xi=1-\frac{a^2}{\ell^2}\,,
\quad \Delta_{\theta}=1-\frac{a^2}{\ell^2}\cos^2\!\theta\ \nonumber\\
\Delta_{r}&=&(r^2+a^2)\Bigl(1+\frac{r^2}{\ell^2}\Bigr)-2Mr+Q^2.
\end{eqnarray}
 In the equatorial plane, the null geodesic condition yields to
\begin{eqnarray}
\gamma_{rr}&=&\frac{r^4}{\left(\left(a^2+r^2\right) \left(\frac{r^2}{\ell^2}+1\right)-2 M r+Q^2\right) \left(r \left(\frac{r \left(a^2+r^2\right)}{\ell^2}-2 M+r\right)+Q^2\right)},\\
\gamma_{\phi\phi}&=&\frac{l^6 r^4 \left(a^2 \left(\ell^2+r^2\right)+\ell^2 \left(-2 M r+Q^2+r^2\right)+r^4\right)}{\left(a^2-\ell^2\right)^2 \left(r^2 \left(a^2+r^2\right)+\ell^2 \left(r (r-2 M)+Q^2\right)\right)^2},\\ \eta_{\phi}&=&-\frac{a \ell^2 \left(r^2 \left(a^2+r^2\right)+\ell^2 \left(Q^2-2 M r\right)\right)}{\left(a^2-\ell^2\right) \left(r^2 \left(a^2+r^2\right)+\ell^2 \left(r (r-2 M)+Q^2\right)\right)}.
\end{eqnarray}
In this model, the  area element on the equatorial plane  is found to be 
\begin{equation}
dS \simeq rdr d\phi+O(M^1,Q^2,a^2,1/\ell^{2}).
\end{equation}
Considering the weak filed and the slow rotation approximations with small values of the  AdS radius, we can obtain the expression of  the Gaussian curvature. Precisely,  it  is given by
\begin{eqnarray}
K&\simeq&\frac{1}{\ell^2}-\frac{6 M}{\ell^2 r}+\frac{6 Q^2}{\ell^2 r^2}-\frac{14 a^2 M}{\ell^2 r^3}-\frac{2 M}{r^3}+\frac{15 a^2 Q^2}{\ell^2 r^4}+\frac{3 Q^2}{r^4}-\frac{6 M Q^2}{r^5}+\frac{24 a^2 M}{\ell^2 r^5}-\frac{6 a^2 M}{r^5}\notag \\&+&\frac{8 a^2 Q^2}{r^6}+\frac{12 a^2 M Q^2}{r^7}+O\left( M^2,Q^3,a^2,\frac{1}{\ell^4}\right).
\end{eqnarray}
Using Eq.(\ref{a37}),
 the integral calculation can be expanded as
 \begin{eqnarray}
- \int^{\phi_{R}}_{\phi_{S}}\int_{r_{o}}^{\infty} K \sqrt{\gamma}drd\phi
&\simeq & \frac{2M}{b}\left[ \sqrt{1-(bu_{S})^{2}}+\sqrt{1-(bu_{R})^{2}}\right]\notag \\&+&\left[ \frac{6 Q^{2}}{\ell^2}-\frac{3Q^{2}}{4b^{2}}\right] \left[\pi  - \arcsin(bu_{S})- \arcsin(bu_{R})\right]\notag \\&-&\frac{3Q^{2}}{4b^{2}}\left[bu_{R}\sqrt{1-(bu_{R})^{2}}+ bu_{S}\sqrt{1-(bu_{S})^{2}} \right] \notag\\&-&\frac{M Q^2}{3 b^3}\left[ \left( 16+(bu_{R})^{2}\right) \sqrt{1-(bu_{R})^{2}}+ \left( 16+(bu_{S})^{2}\right) \sqrt{1-(bu_{S})^{2}}\right] \notag \\&+& \frac{b}{2 \ell^2}\left[ \frac{\sqrt{1-(b u_{R})^2}}{u_{R}}+\frac{\sqrt{1-(b u_{S})^2}}{u_{S}}\right]\notag \\ &-&\frac{M b}{2\ell^2}\left[ \frac{1}{\sqrt{1-(b u_{S})^2}}+\frac{1}{\sqrt{1-(b u_{R})^2}}\right] \\\
&-&\frac{6 Q^{2}b}{\ell^2}\left[ u_{S}\arctan(u_{S}b)+u_{R}\arctan(u_{R}b))\right] +O\left( M^2,Q^3,a,\frac{1}{\ell^4}\right). \notag
\label{kk}
 \end{eqnarray}
To get the expression of $k_{g}$, one should exploit Eq(\ref{a28}). Indeed,    the  computations of the geodesic curvature give
\begin{eqnarray}
k_{g}&\simeq&-\frac{2a}{\ell^2}-\frac{a M}{\ell^2 r}-\frac{2 a M}{r^3}-\frac{3 a M Q^2}{2 \ell^2 r^3}+\frac{2 a Q^2}{r^4}+\frac{3 a M Q^2}{r^5}+O\left( M^2,Q^3,a,\frac{1}{\ell^4}\right).
\label{a48}
\end{eqnarray}
For the prograde  case  where $dl>0$, we consider  $r=b/ \cos \vartheta$ and $l= b \tan\vartheta$. This linear approximation of the photon  is needed to  get the integratio computations.  In particular, we obtain
\begin{eqnarray}
\int^{R}_{S}k_{g}dl&\simeq&\int^{R}_{S}-\frac{2a b}{\ell^2\cos^{2}\vartheta}-\frac{a M}{\ell^2\cos\vartheta}-\left[ \frac{2 a M}{b^2}+\frac{3 a M Q^2}{2 \ell^2 b^2}\right] \cos\vartheta+\frac{2 a Q^2}{b^3}\cos^{2}\vartheta+\frac{3 a M Q^2}{b^4}\cos^{3}\vartheta d\vartheta \notag \\
&=& -\frac{2a}{\ell^2}\left[ \frac{\sqrt{1-(b u_{R})^2}}{u_{R}}+\frac{\sqrt{1-(b u_{S})^2}}{u_{S}}\right]+ \frac{a Q^2}{b^{3}}\left[\pi- \arcsin(bu_{S})-\arcsin(bu_{R})\right] \notag \\&-&\frac{a M}{\ell^2 } \left[ \pi-\arctan(\sqrt{1-(u_{S}b)^2})-\arctan(\sqrt{1-(u_{R}b)^2})\right]\notag\\ &-&\left[ \frac{2 a M}{b^2}+\frac{3 a M Q^2}{2 \ell^2 b^2}-\frac{2 a M Q^2}{b^4}\right]\left[ \sqrt{1-(bu_{S})^{2}}+\sqrt{1-(bu_{R})^{2}}\right]\notag \\&+&\frac{a Q^2}{b^2}\left[ u_{S}\sqrt{1-(u_{S}b)^2}+u_{R}\sqrt{1-(u_{R}b)^2}\right] \notag \\&+&\frac{a MQ^2}{b^2}\left[ u_{S}^2\sqrt{1-(u_{S}b)^2}+u_{R}^2\sqrt{1-(u_{R}b)^2}\right] +O\left( M^2,Q^3,a,\frac{1}{\ell^4}\right).
\label{kkg}
\end{eqnarray}
Combining  Eqs.(\ref{kk}) and (\ref{kkg}),
the limits $bu_{S}<< 1$ and  $bu_{R}<<1$ provide
\begin{eqnarray}
\Theta_{KN_{AdS}} &\simeq& \frac{4M}{b}-\frac{3Q^2 \pi}{4 b^2}-\frac{4 a M}{b^2}-\frac{3aM Q^2}{\ell^2 b^2}-\frac{32M Q^2}{3 b^3}+\frac{a Q^2 \pi}{b^3}+\frac{4 a M Q^2}{b^4}+\frac{6 Q^2 \pi}{\ell^2}-\frac{aM \pi}{2 \ell^2}\notag \\&-&\frac{Mb}{\ell^2}+\left[ \frac{b}{2\ell^2}-\frac{2 a}{\ell^2}\right] \left[\frac{1}{u_{R}}+\frac{1}{u_{S}} \right] +O\left( M^2,Q^3,a,\frac{1}{\ell^4}\right).
\label{thetak}
\end{eqnarray}
From Eq.(\ref{thetak}), we see that we can
recover the rotating  and the  charge  contribution terms obtained in the Kerr-Newman  ordinary black holes \cite{O}. Considering gravitational bending angle of lights for finite distances where the observer and  the source location as $u_{S}=u_{R}=0.1$, we illustrate the variation of the deflection angle $\Theta$ in terms of  the impact parameter by varying the rotating parameter for fixed values of the mass, the charge  and the AdS radius. These  behaviors are  plotted in Fig(\ref{im5}). 

To explore the  effect of the cosmological constant, we consider two specific AdS radius values. Indeed, we examine $\ell=\infty$ and   $\ell=20$ presented in  left and right panels, respectively. In the absence of AdS  backgrounds, the deflection angle decreases by increasing the rotating parameter $a$. It remains  decreasing for generic values of $b$.  The right panel of the figure shows that the deflection angle decreases rapidly for small values of $b$  then  it becomes an increasing function. Taking  a generic value of $\ell$,   the deflection angle decreases with the rotation parameter.    In the fixed charge solutions, however,   the previous  behavior changing point is not observed for the  rotation parameter variation.  It is  worth noting  that, in this case, the  variation of the charge   provides a   critical point $b^ a_c$ where the charge effect is  inverted.  This critical point can be also obtained by   using  the deflection angle  charge derivative.  This value should    depend  on  the three parameters $\ell$, $M$ and $a$.  Indeed,  the computation gives 
\begin{equation}
b^a_{c}=b_{c}+a \left(\frac{\mu+ \nu}{(6 \pi )^{2/3} \lambda ^{4/3} \left(3 \pi ^2 \ell ^2-32768 M^2\right)}\right)
\end{equation}
where one has used 
\begin{eqnarray}
\mu&=& -16777216 \sqrt[3]{6} M^4 \ell ^2-65536 \sqrt[3]{6} M^3 \left(\lambda -256 \text{M$\ell $}^2\right)+512 M^2 \left(256 \pi ^{4/3} \lambda ^{2/3} \ell ^2-253 \sqrt[3]{6} \pi ^2 \ell ^4\right)  \nonumber \\
\nu&=&  \pi ^{4/3} M \left(\lambda -256 \text{M$\ell $}^2\right) \left(515 \lambda ^{2/3}-509 \sqrt[3]{6} \pi ^{2/3} \ell ^2\right)+12 \sqrt[3]{6} \pi ^4 \ell ^6 \\&-&12 \pi ^{10/3} \lambda ^{2/3} \ell ^4}{(6 \pi )^{2/3} \lambda ^{4/3} \left(3 \pi ^2 \ell ^2-32768 M^2\right).\nonumber
\end {eqnarray}
Sending $a$ to zero, we recover the previous relation associated with non rotating solutions. 
\section{Plasma medium and deflection angle of light rays}
\begin{figure}[!ht]
		\begin{center}
		\begin{tikzpicture}[scale=0.2,text centered]
\node[] at (-45,1){\small  \includegraphics[scale=0.6]{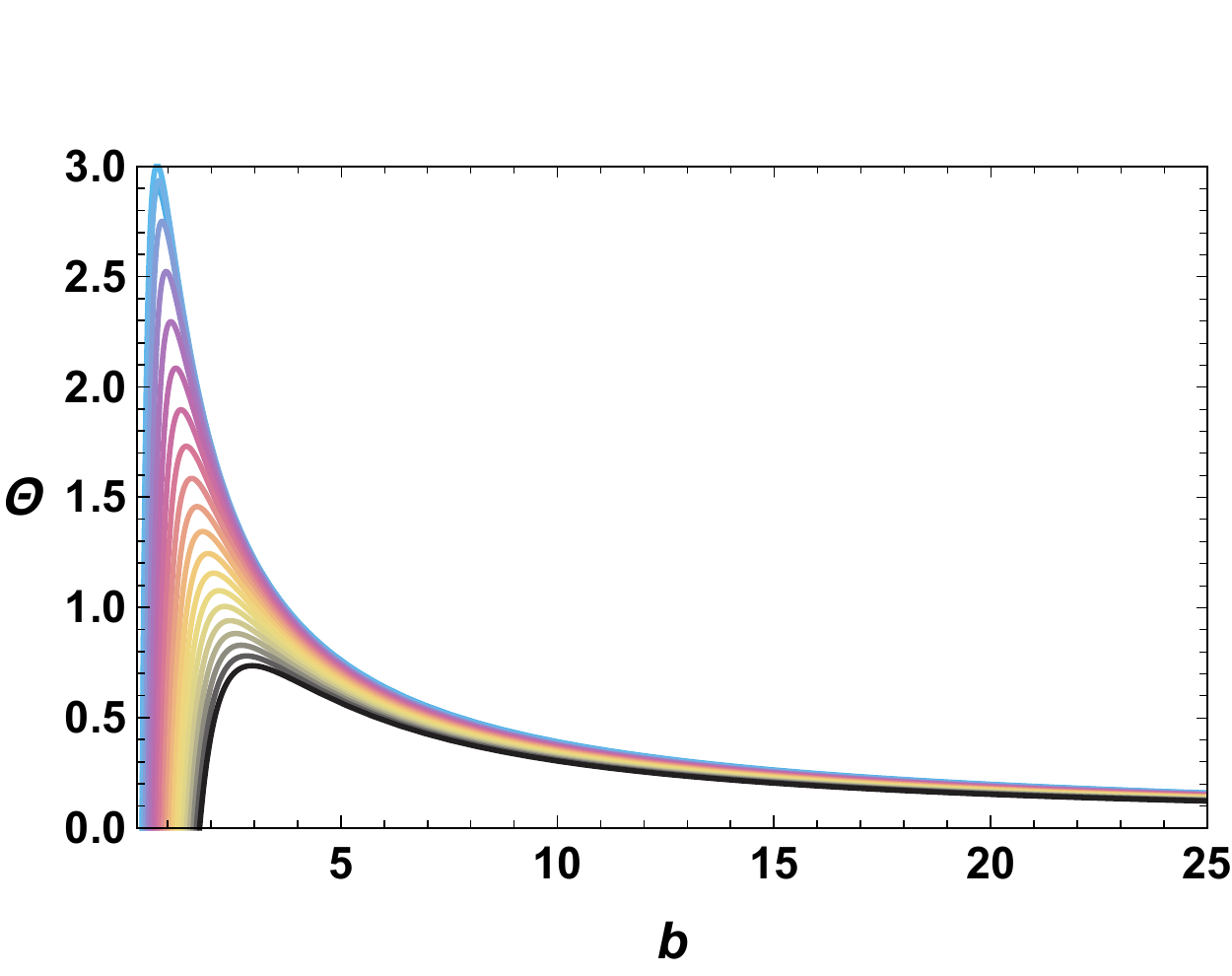}};	
\node[] at (0,1){\small  \includegraphics[scale=0.6]{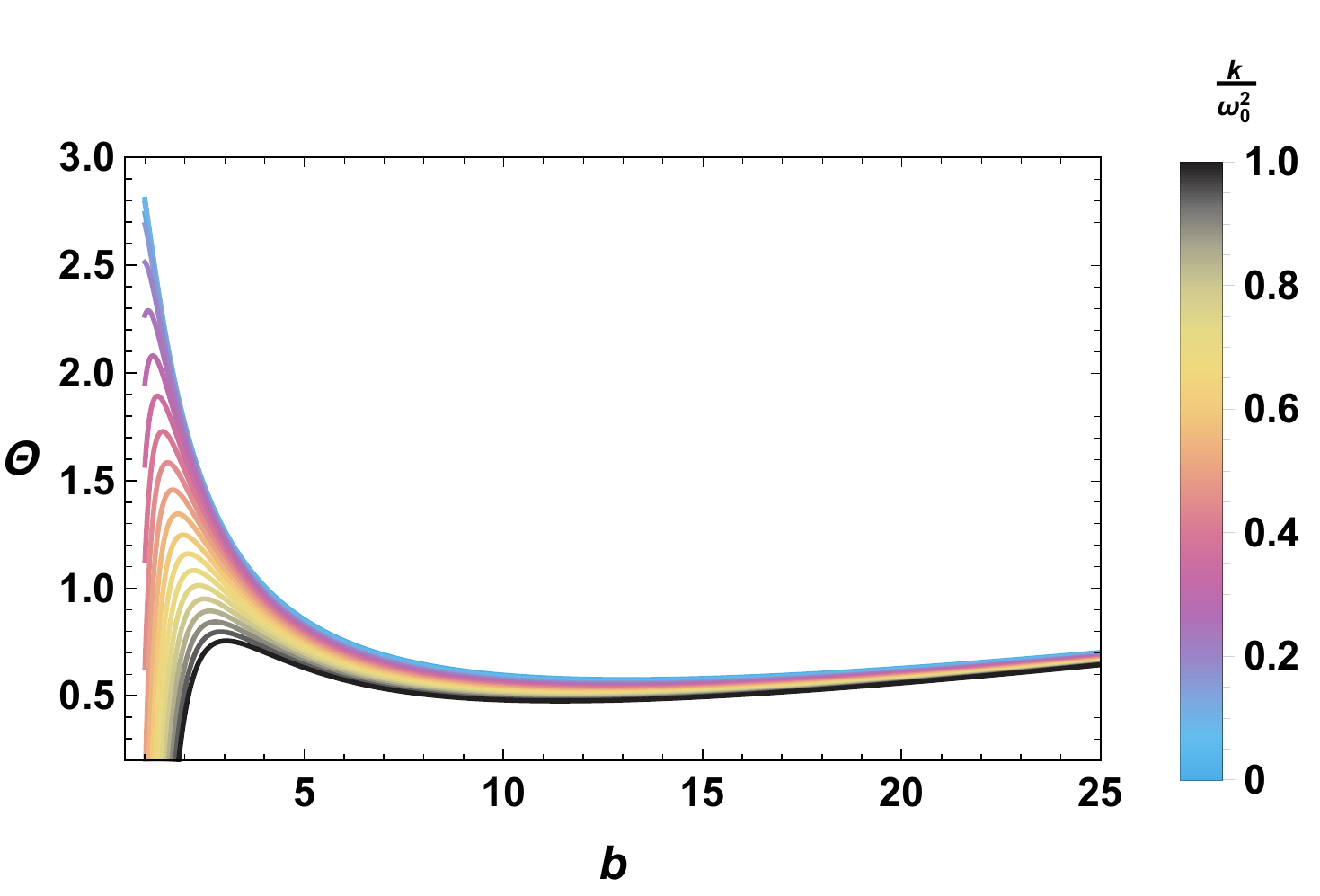}};	
\node[] at (3,3){\small  \includegraphics[scale=0.3]{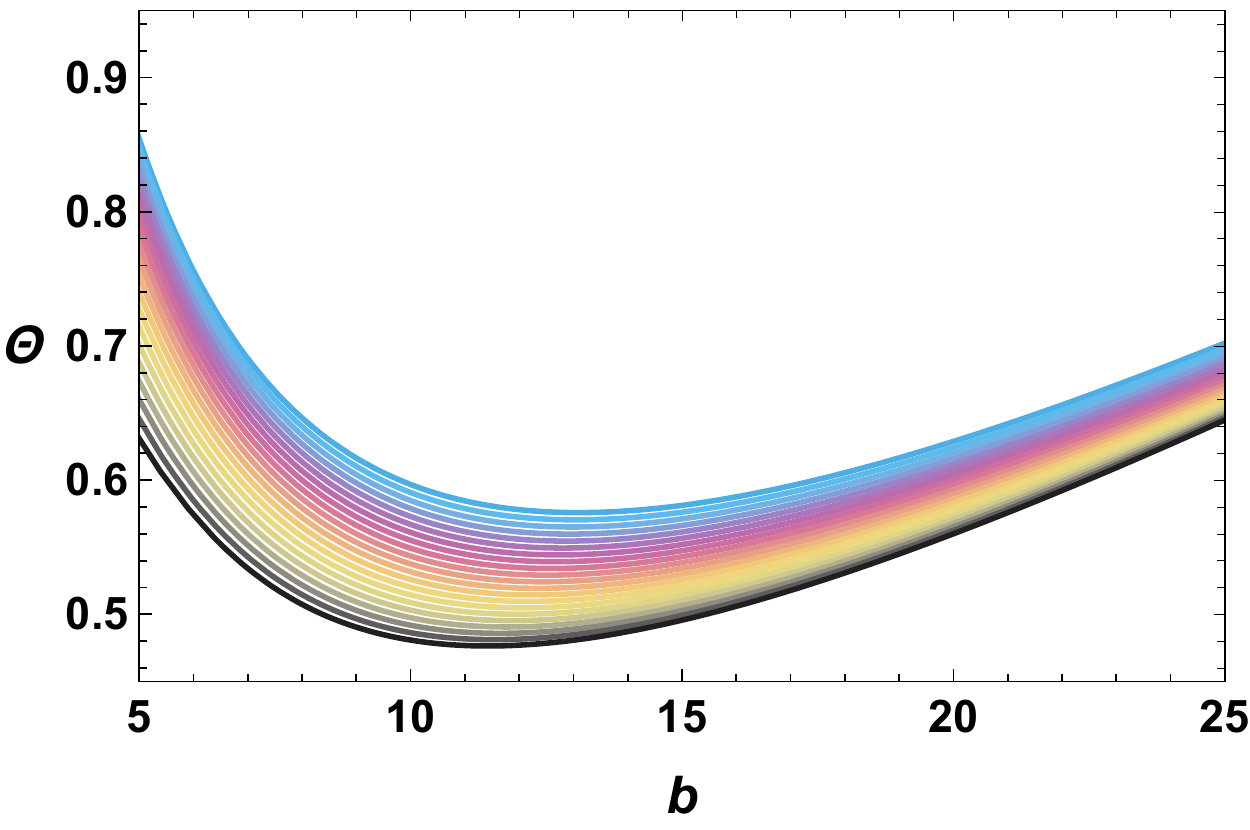}};	
\node[] at (-39,3){\small  \includegraphics[scale=0.3]{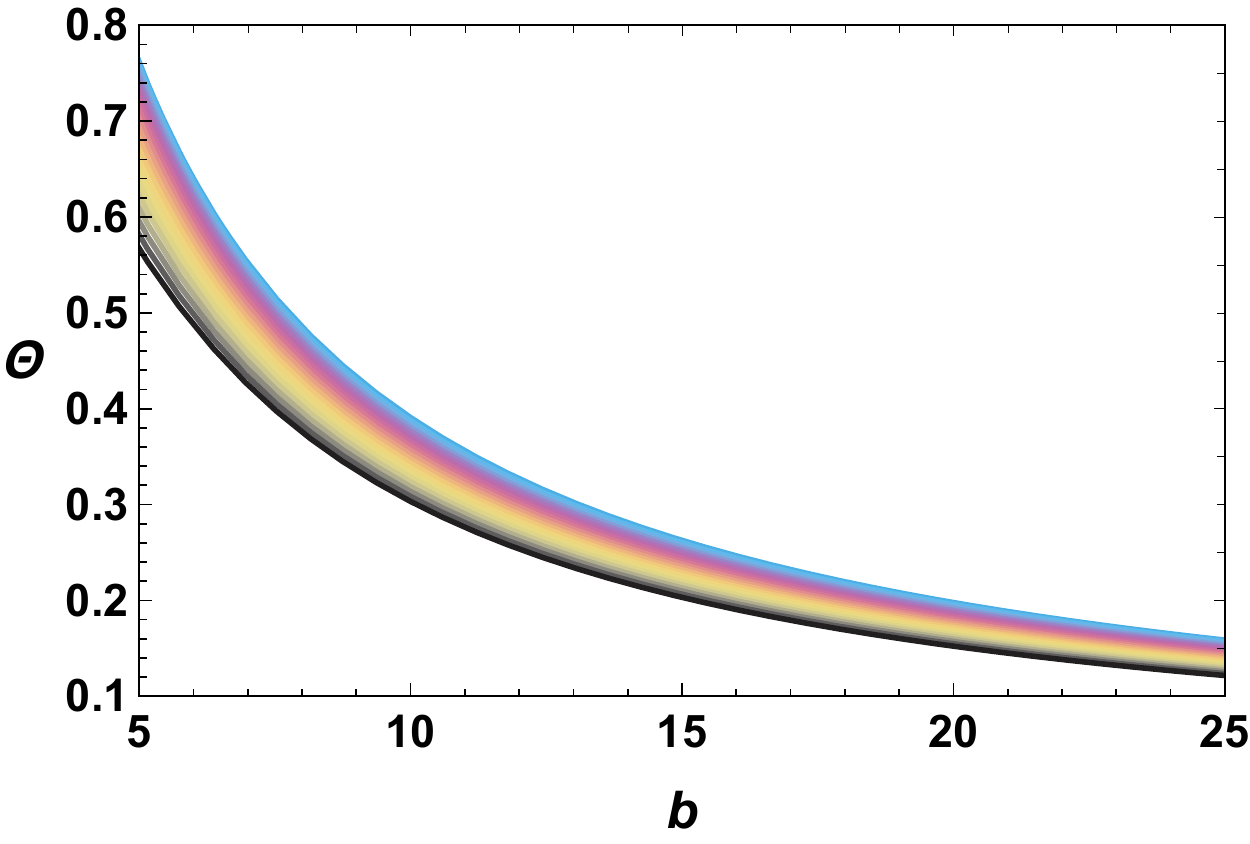}};	
\end{tikzpicture}	
\caption{Variation of the deflection angle in terms of the frequency ratio $\frac{k}{\omega_{0}^{2}}$ for $u_{S}=u_{R}=0.1$, $Q=0.2$, $a=0.2$ and $M=1$. Left panel: behavior with $\ell=\infty$, right panel: behavior with $\ell=20$.}
\label{fn}
\end{center}
\end{figure}

In this section, we investigate the plasma effect on the weak  deflection angle of four-dimensional  AdS black holes. For simplicity reasons, we consider only the  Kerr-Newman AdS  solution by assuming that the calculations for other models can be done in a similar way. To do so, certain optical quantities are needed.  According to \cite{sss}, the refraction  index in terms of the photon four-momentum  and velocity $u^{B}$   can be written as
\begin{equation}
n^{2}=1+\frac{P_{\alpha}P^{\alpha}}{(P_{B}u^{B})^2}.
 \label{nr}
\end{equation}
For a non-magnetized cold plasma,  Eq.(\ref{nr}) reduces to the
following form
\begin{equation}
n^{2}=1-\frac{\omega _{p}(r)^{2}}{\omega_{0}^2},
\end{equation}
where $\omega_{0}^2$  represents  the photon frequency and $\omega _{p}(p)$
 is the electron plasma frequency.    A priory,  they are  many  models which  have been dealt with. Here, however, we consider  a   specific form studied in  \cite{54,56}.   Using a radial power law, the associated photon  frequency   can be written as follows
 \begin{equation}
\omega^2_{p}(r)=\frac{k}{r^h}, \qquad h>0,
\end{equation}
where $k$ is an arbitrary constant being related to certain physical quantities.   To inspect  such  an effect, we  examine a concrete model relying on an inhomogeneous space-time. For the sake of simplicity, we take $h=1$. In this way, the two dimensional  optical geometry of the Kerr-Newman AdS black hole, in the presence of the plasma medium,  is given by
\begin{equation}
dl=n^2 \gamma_{ij}dx^{i}dx^{j},
\end{equation}
being a generalized form of Eq.(\ref{labe}). This  provides the  following expression of the Gaussian optical curvature
 \begin{eqnarray}
 K&=& \frac{1}{\ell ^2}+\left( -\frac{6 M}{\ell ^2}+\frac{3 k}{2 \omega _0^2 \ell ^2}\right) \frac{1}{r}+\left(\frac{6 Q^2}{\ell ^2}+ \frac{3 k^2}{\omega _0^4 \ell ^2}-\frac{15 k M}{2 \omega _0 ^{2}\ell ^2}\right) \frac{1}{r^2}\notag \\&+&\left(-2 M+ \frac{11 k^3}{2 \omega _0^6 \ell ^2}-\frac{11 k^2 M}{\omega _0^4 \ell ^2}+\frac{7 k Q^2}{\omega _0^2 \ell ^2}+\frac{k}{2 \omega _0^2}\right) \frac{1}{r^3}\notag \\&+&\left(3 Q^2+ \frac{9 k^4}{\omega _0^8 \ell ^2}-\frac{33 k^3 M}{2 \omega _0^6 \ell ^2}+\frac{9 k^2 Q^2}{\omega _0^4 \ell ^2}+\frac{3 k^2}{2 \omega _0^4}-\frac{9 k M}{2 \omega _0^2}\right) \frac{1}{r^4} \\&+&\left(-6 M Q^2+ \frac{27 k^5}{2 \omega _0^{10} \ell ^2}-\frac{24 k^4 M}{\omega _0^8 \ell ^2}+\frac{12 k^3 Q^2}{\omega _0^6 \ell ^2}+\frac{3 k^3}{\omega _0^6}-\frac{9 k^2 M}{\omega _0^4}+\frac{9 k Q^2}{2 \omega _0^2}\right) \frac{1}{r^5}+O\left( M^2,Q^3,a,\frac{1}{\ell^2}\right)\notag .
 \end{eqnarray}
The first part of the deflection angle equation could be expended as follows
 \begin{eqnarray}
 - \int^{\phi_{R}}_{\phi_{S}}\int_{r_{o}}^{\infty} K \sqrt{\gamma}drd\phi
&= & \left(2 M- \frac{11 k^3}{2 \omega _0^6 \ell ^2 }+\frac{11 k^2 M}{\omega _0^4 \ell ^2 }-\frac{7 k Q^2}{\omega _0^2 \ell ^2 }-\frac{k}{2 \omega _0^2 }\right)\left(\frac{\sqrt{1-(bu_{S})^{2}}+\sqrt{1-(bu_{R})^{2}}}{b} \right) \notag
\end{eqnarray}
\begin{eqnarray}
&-&\left(\frac{3 Q^2}{4 b^2}+ \frac{9 k^4}{\omega _0^8 \ell ^2 4 b^2}-\frac{33 k^3 M}{8 \omega _0^6 \ell ^2  b^2}+\frac{9 k^2 Q^2}{4 \omega _0^4 \ell ^2  b^2}+\frac{3 k^2}{8 \omega _0^4  b^2}-\frac{9 k M}{8 \omega _0^2  b^2}\right)\notag\\
&\times&\left(\left[\pi  - \arcsin(bu_{S})- \arcsin(bu_{R})\right]+ \left[bu_{R}\sqrt{1-(bu_{R})^{2}}+ bu_{S}\sqrt{1-(bu_{S})^{2}} \right]\right)\notag \\&+&\left(-6 M Q^2+ \frac{27 k^5}{2 \omega _0^{10} \ell ^2}-\frac{24 k^4 M}{\omega _0^8 \ell ^2}+\frac{12 k^3 Q^2}{\omega _0^6 \ell ^2}+\frac{3 k^3}{\omega _0^6}-\frac{9 k^2 M}{\omega _0^4}+\frac{9 k Q^2}{2 \omega _0^2}\right)\frac{1}{18 b^3}\notag\\
&\times&  \left[ \left( 16+(bu_{R})^{2}\right) \sqrt{1-(bu_{R})^{2}}+ \left( 16+(bu_{S})^{2}\right) \sqrt{1-(bu_{S})^{2}}\right]\notag \\&+& \frac{b}{2 \ell^2}\left( \frac{\sqrt{1-(b u_{R})^2}}{u_{R}}+\frac{\sqrt{1-(b u_{S})^2}}{u_{S}}\right)+\left( -\frac{ M b}{ 2\ell ^2}+\frac{ k b}{8 \omega _0^2 \ell ^2}\right)\left( \frac{1}{\sqrt{1-(b u_{S})^2}}+\frac{1}{\sqrt{1-(b u_{R})^2}}\right)\notag \\&+& \left(\frac{6 Q^2}{\ell ^2}+ \frac{3 k^2}{\omega _0^4 \ell ^2}-\frac{15 k M}{2 \omega _0^{2} \ell ^2}\right)\left( \left[\pi  - \arcsin(bu_{S})- \arcsin(bu_{R})\right]-b\left[ u_{S}\arctan(u_{S}b)+u_{R}\arctan(u_{R}b)\right] \right) \notag \\&+& O\left( M^2,Q^3,a,\frac{1}{\ell^2}\right).
\label{Kn}
 \end{eqnarray}
Moreover,  the  geodesic curvature is found to be
 \begin{eqnarray}
 k_{g}&=&-\frac{2 a}{\ell ^2}+\left( -\frac{2 a k}{\omega _0 ^{2}\ell ^2}-\frac{a M}{\ell ^2}\right) \frac{1}{r}-\left(\frac{2 a k^2}{\omega _0^4 \ell ^2}+\frac{a k M}{\omega _0 ^{2}\ell ^2} \right)\frac{1}{r^2}+\left(-2 a M
  -\frac{2 a k^3}{\omega _0^6 \ell ^2}-\frac{a k^2 M}{\omega _0^4\ell ^2}-\frac{3 a M Q^2}{2 \ell ^2}\right) \frac{1}{r^3}\notag \\&+&\left(2a Q^2-\frac{2 a k^4}{\omega _0^8 \ell ^2}-\frac{a k^3 M}{\omega _0^6 \ell ^2}-\frac{3 a k M Q^2}{2 \omega _0^{2} \ell ^2}-\frac{2 a k M}{\omega _0^{2}} \right) \frac{1}{r^4}\notag\\&+&\left(3 a M Q^2-\frac{2 a k^5}{\omega _0^{10} \ell ^2}-\frac{a k^4 M}{\omega _0^8\ell ^2}-\frac{3 a k^2 M Q^2}{2 \omega _0^4 \ell ^2}-\frac{2 a k^2 M}{\omega _0^4}+\frac{2 a k Q^2}{\omega _0^{2}} \right) \frac{1}{r^5} \notag \\&+& O\left( M^2,Q^3,a,\frac{1}{\ell^2}\right).
 \end{eqnarray}
Integrating  this expression, we obtain
\begin{eqnarray}
 \int^{R}_{S}k_{g}dl
&=& -\frac{2a}{\ell^2}\left( \frac{\sqrt{1-(b u_{R})^2}}{u_{R}}+\frac{\sqrt{1-(b u_{S})^2}}{u_{S}}\right) \notag \\&+&\left( -\frac{2 a k}{\omega _0^{2} \ell ^2}-\frac{a M}{\ell ^2}\right)\left(\pi-\arctan(\sqrt{1-(u_{S}b)^2})-\arctan(\sqrt{1-(u_{R}b)^2})\right)\notag\\
&-&\left(\frac{2 a k^2}{\omega _0^4 \ell ^2}+\frac{a k M}{\omega _0 ^{2}\ell ^2} \right)\left(\frac{\pi- \arcsin(bu_{S})+\arcsin(bu_{R})}{b}\right)\notag\\&+& \left(-2 a M-\frac{2 a k^3}{\omega _0^6 \ell ^2}-\frac{a k^2 M}{\omega _0^4 \ell ^2}-\frac{3 a M Q^2}{2 \ell ^2}\right)\left(\frac{\sqrt{1-(bu_{S})^{2}}+\sqrt{1-(bu_{R})^{2}}}{b^2} \right)\notag\\&+&\left(2 a Q^2-\frac{2 a k^4}{\omega _0^8 \ell ^2}-\frac{a k^3 M}{\omega _0^6 \ell ^2}-\frac{3 a k M Q^2}{2 \omega _0^{2} \ell ^2}-\frac{2 a k M}{\omega _0^{2}} \right)\notag\\&\times& \left( \frac{ u_{S}\sqrt{1-(u_{S}b)^2}+u_{R}\sqrt{1-(u_{R}b)^2}}{2b^2} +\frac{\pi- \arcsin(bu_{S})+\arcsin(bu_{R})}{2b^3}\right) \notag\\
 &+&\left(3 a M Q^2-\frac{2 a k^5}{\omega _0^{10} \ell ^2}-\frac{a k^4 M}{\omega _0^8 \ell ^2}-\frac{3 a k^2 M Q^2}{2 \omega _0^4 \ell ^2}-\frac{2 a k^2 M}{\omega _0^4}+\frac{2 a k Q^2}{\omega _0^{2}} \right) \notag\\ &\times & \left(\frac{u_{S}^2\sqrt{1-(u_{S}b)^2}+u_{R}^2\sqrt{1-(u_{R}b)^2}}{3b^2}+ \frac{2\left(\sqrt{1-(bu_{S})^{2}}+\sqrt{1-(bu_{R})^{2}} \right) }{3 b^4} \right) \notag \\&+& O\left( M^2,Q^3,a,\frac{1}{\ell^2}\right).
 \label{kgn}
  \end{eqnarray}
Combining Eq.(\ref{Kn}) and Eq.(\ref{kgn}) in linear approximations, we can get   the deflection angle of the  Kerr AdS black hole in such a  medium.  Taking $bu_{S}<< 1$ and  $bu_{R}<<1$, we obtain  this optical quantity
 \begin{eqnarray}
\Theta &=&  \Theta_{KN_{AdS}} -\frac{8 a k^5}{3 b^4 \omega _0^{10} \ell ^2}-\frac{4 a k^4 M}{3 b^4 \omega _0^8 \ell ^2}-\frac{2 a k^2 M Q^2}{b^4 \omega _0^4 \ell ^2}-\frac{8 a k^2 M}{3 b^4 \omega _0^4}+\frac{8 a k Q^2}{3 b^4 \omega _0^2}-\frac{\pi  a k^4}{b^3 \omega _0^8 \ell ^2}-\frac{\pi  a k^3 M}{2 b^3 \omega _0^6 \ell ^2}\notag \\&-&\frac{3 \pi  a k M Q^2}{4 b^3 \omega _0^2 \ell ^2} -\frac{\pi  a k M}{b^3 \omega _0^2}-\frac{4 a k^3}{b^2 \omega _0^6 \ell ^2}-\frac{2 a k^2 M}{b^2 \omega _0^4 \ell ^2}-\frac{2 \pi  a k^2}{b \omega _0^4 \ell ^2}-\frac{\pi  a k M}{b \omega _0^2 \ell ^2}-\frac{\pi  a k}{\omega _0^2 \ell ^2}+\frac{24 k^5}{b^3 \omega _0^{10} \ell ^2}-\frac{128 k^4 M}{3 b^3 \omega _0^8 \ell ^2}\notag \\&+&\frac{64 k^3 Q^2}{3 b^3 \omega _0^6 \ell ^2}+\frac{16 k^3}{3 b^3 \omega _0^6}-\frac{16 k^2 M}{b^3 \omega _0^4}+\frac{8 k Q^2}{b^3 \omega _0^2}-\frac{9 \pi  k^4}{4 b^2 \omega _0^8 \ell ^2}+\frac{33 \pi  k^3 M}{8 b^2 \omega _0^6 \ell ^2}-\frac{9 \pi  k^2 Q^2}{4 b^2 \omega _0^4 \ell ^2}-\frac{3 \pi  k^2}{8 b^2 \omega _0^4}+\frac{9 \pi  k M}{8 b^2 \omega _0^2}\notag \\&-&\frac{11 k^3}{b \omega _0^6 \ell ^2}+\frac{22 k^2 M}{b \omega _0^4 \ell ^2}-\frac{14 k Q^2}{b \omega _0^2 \ell ^2}-\frac{k}{b \omega _0^2}+\frac{b k}{4 \omega _0^2 \ell ^2}+\frac{3 \pi  k^2}{\omega _0^4 \ell ^2}-\frac{15 \pi  k M}{2 \omega _0^2 \ell ^2} +O\left( M^2,Q^3,a,\frac{1}{\ell^2}\right).
\end{eqnarray}
At this level, we could provide certain comments.   Taking $k=0$, we recover the result of the ordinary solutions $\Theta=\Theta_{KN_{AdS}}$.
 As usually,  it  follows  that    the deflection angle expression involves  a  refractive index contribution. To investigate the  effect of the  plasma medium  on  the deflection angle of RN-AdS black holes, we consider as previously   two AdS radius values. In particular, we vary the following frequency ratio $\frac{k}{\omega_{0}^{2}}$ from 0 to 1.  Indeed,    Fig(\ref{fn}) illustrates such  behaviors. The  left and the right panels of this figure show that the deflection angle  of light rays remains a decreasing function in terms of the fraction frequency quantity by  preserving the same contribution even with large values of $b$.   In the right panel where the AdS space-time contribution is visualized by taking $\ell=20$, the deflection angle increases for large impact parameter values. This behavior could be explained by the presence of the cosmological constant.   This can affect  certain parameters as  the charge. However, the rotation parameter and the  frequency ratio  keep the same behaviors for large and small values of the impact parameter in the presence of the AdS space-time.
 \section{Conclusions}
In this work, we have investigated  the deflection angle of AdS black holes in four dimensions. In particular, we have  examined the dependence of   such a quantity in  terms  of black hole parameters.   Presenting  the needed formalism  using the Gauss-Bonnet theorem,  we  have first  computed the deflection angle of RN-AdS black holes. In such a model,   we have  shown that  the deflection angle is a decreasing function of the charge  and the AdS radius.   It has been observed  that the deflection angle involves a maximum. This maximum   has been increasing by decreasing the charge $Q$.  In particular, we  have found  a specific point  where  the  behavior of the deflection angle in terms of the charge  changes from a decreasing function  to an increasing one.  After that,   we  have studied   optical aspects   of  the  Kerr-Newman AdS black holes using the deflection angle variations.
Precisely,   we have remarked that the behavior  of the deflection 	angle, as a  decreasing function of the rotating  parameter,  does not change in the presence of the AdS space. \\
Motivated by observational aspects  of the plasma influence on the deflection angle being studied by exploiting several astrophysical scenarios,   we have inspected the plasma medium  effect.  Concretely, we have considered   the variation of the  frequency ratio.  Among others,  we have remarked that the deflection angle  is an  increasing function in terms  of the ratio $\frac{k}{\omega_{0}^{2}}$.  At this level, we  have  observed that, in    the plasma medium presence,  the deflection angle  keeps  the same behavior  appearing in the rotating black hole  with AdS  geometries.\\
 This work comes  up with certain open questions. Higher dimensional models could be  a possible extension of the present work. Dark energy and dark matter effects could be also   implemented. We hope address such questions  in future works.\\
  {\bf Data availability}\\
  Data sharing is not applicable to this article. 
\section*{Acknowledgments}
The authors would like to thank H. El Moumni, Y. Hassouni and M.  B. Sedra for collaborations on related topics.
They   are also  grateful to the
anonymous referee for insightful comments and
suggestions. 
 This work is partially
supported by the ICTP through AF.


\begin{thebibliography}{10}
\bibitem{emparan}
 R. Emparan, H. S. Reall, \textit{Black Holes in Higher Dimensions},  Living Rev. Relativity, {\bf 11}(1) (2008) 6, {\tt arXiv:0801.3471}.

\bibitem{SWH}
S. W. Hawking, \textit{Black hole explosions}, Nature {\bf 248} (1974) 30.
\bibitem{sss}
J. L. Synge, \textit{Relativity: The General Theory}. North Holland, Amsterdam (1960).
\bibitem{A1}
 B. Abbott and al., \textit{Observation of Gravitational Waves from a Binary Black Hole Merger},
Phys.\ Rev.\ Lett.\ {\bf 116} (6) (2016) 061102, {\tt arXiv:1602.03837}.
\bibitem{A2}
K. Akiyama and al.,
\textit{First M87 Event Horizon Telescope Results. IV. Imaging the Central
  Supermassive Black Hole},
 Astrophys. J. {\bf L4} (1) (2019) 875, {\tt arXiv:1906.11241}.

\bibitem{A3}
K. Akiyama and al.,
\textit{First M87 Event Horizon Telescope Results. V. Imaging the Central
  Supermassive Black Hole},
 Astrophys. J. {\bf L5} (1) (2019) 875.
\bibitem{A4}
K. Akiyama and al.,
\textit{First M87 Event Horizon Telescope Results. VI. Imaging the Central
  Supermassive Black Hole},
 Astrophys. J. {\bf L6} (1) (2019) 875.
 \bibitem{E1} Y. Liu, D. C. Zou, B. Wang, \textit{Signature of the Van der Waals like small-large charged AdS black hole phase transition in quasinormal modes}, JHEP. \textbf{09} (2014) 179, {\tt arXiv:1405.2644}.


\bibitem{F} A. Belhaj, M. Chabab, H. El Moumni, K. Masmar, M. B. Sedra, A. Segui, \textit{On heat properties of AdS black holes in higher dimensions}, JHEP. \textbf{05} (2015) 149, {\tt arXiv:1503.07308}.
\bibitem{F1}
 A. Belhaj, M. Chabab, H. El Moumni, L. Medari, M. B. Sedra, \textit{The thermodynamical behaviors of Kerr–Newman AdS black holes}, CPL. \textbf{30}  (2013)  090402, {\tt arXiv:1307.7421}.
 \bibitem{12}
D. Kubizňák, R. B. Mann, Mae Teo, \textit{Black hole chemistry: thermodynamics
with Lambda}, Class. Quantum Grav. \textbf{34} (2017) 063001, {\tt arXiv:1608.06147}.
\bibitem{f2}
 A. Rajagopal,  D. Kubiznak, R. B. Mann,  \textit{Van der Waals black hole}, Phys. Lett. B \textbf{737} (2014) 277, {\tt arXiv:1408.1105}.
  \bibitem{D}
  S. W. Hawking, D. N. Page, \textit{Thermodynamics of black holes in anti-de Sitter space}, Commun. Math. Phys. \textbf{87} (4) (1983) 577.
  \bibitem{7}
Y-Y. Wang, B-Y Su, N. Li, \textit{Hawking Page phase transitions in four-dimensional Einstein Gauss-Bonnet gravity}, Phys. Dark Universe \textbf{31} (2021) 100769, {\tt arXiv:2008.01985}.
\bibitem{14}
H. Quevedo, A. Sanchez, S. Taj, A. Vazquez, \textit{Phase transitions in geometrothermodynamics}, Gen. Rel. Grav \textbf{43}  (2011) 1153, {\tt arXiv:1010.5599}.
\bibitem{E}
   A. Belhaj, M. Chabab, H. El Moumni, M. B. Sedra, \textit{On thermodynamics of AdS black holes in arbitrary dimensions},
 CPL. \textbf{29} (2012) 100401, {\tt arXiv:1210.4617}.
 \bibitem{Carlo}
P.~V.~Cunha, C.~A.~R.~Herdeiro, B.~Kleihaus, J.~Kunz,  E.~Radu,
\textit{Shadows of Einstein–dilaton–Gauss–Bonnet black holes},
Phys. Lett. B\textbf{768}(2017)373, 
{\tt arXiv:1701.00079}.
\bibitem{G}
A.~Övgün, I.~Sakalli, J.~Saavedra,
\textit{Shadow cast and Deflection angle of Kerr-Newman-Kasuya spacetime},
JCAP. \textbf{10} (2018) 041, {\tt arXiv:1807.00388}.
\bibitem{H}
 A. Belhaj, H. Belmahi,  M. Benali, W. El Hadri, H. El Moumni, E. Torrente-Lujan,  \textit{Shadows of 5D Black Holes from string theory},  Phys. Lett.
B \textbf{812} (2021) 136025, {\tt arXiv:2008.13478}.
\bibitem{I}
R. Konoplya, \textit{Shadow of a black hole surrounded by dark matter}, Phys. Lett.
B \textbf{795} (2019) 1, {\tt arXiv:1905.00064}.
\bibitem{J}
S. U. Khan, J. Ren, \textit{Shadow cast by a rotating charged black hole in quintessential dark energy}, Phys. Dark Univ. \textbf{30} (2020) 100644, {\tt arXiv:2006.11289}.
\bibitem{RCDM}
X. Hou, Z. Xu, J. Wang, \textit{Rotating black hole shadow in perfect
fluid dark matter}, JCAP \textbf{12} (2018) 040.
\bibitem{K}
S. W. Wei, Y. C. Zou, Y. X. Liu, R. B. Mann, \textit{Curvature radius and Kerr black hole
shadow}, JCAP \textbf{08} (2019) 030, {\tt arXiv:1904.07710}.
\bibitem{4}
W. Javed, J. Abbas, A.~Övgün, \textit{ Effect of the quintessential dark energy on weak
deflection angle by Kerr Newmann Black hole}, Annals of Physics \textbf{418}  (2020) 168183, {\tt arXiv:2007.16027}.
\bibitem{M}
J. R. Villanueva, J. Saavedra, M. Olivares, N. Cruz, \textit{Photons motion in charged Anti-de Sitter black holes}, Astrophys. Space Sci. \textbf{344} (2) (2013) 437.
\bibitem{DA1}
 W. Javed,  M. B. Khadim, J. Abbas,  A.Övgün, \textit{Weak gravitational lensing by stringy black holes}, Eur. Phys. J. Plus \textbf{3} (2020) 135, {\tt arXiv:2004.00408 }
 \bibitem{5}
A. Belhaj, M. Benali, A. El Balali, H. El Moumni, S-E. Ennadifi,
\textit{Deflection angle and shadow behaviors of quintessential black holes in arbitrary dimensions}, Class. Quantum Grav. \textbf{37} (2020) 215004, {\tt arXiv:2006.01078}.
 \bibitem{1}
R. Uniyal, H. Nandan, P. Jetzer, \textit{Bending angle of light in equatorial plane of Kerr–Sen Black Hole}, Phys. Lett. B  \textbf{782} (2018) 185, {\tt arXiv:1803.04268}.
\bibitem{L}
G. W. Gibbons, M. C. Werner, \textit{Applications of the Gauuss–Bonnet theorem to
gravitational lensing}, Class. Quantum Grav. \textbf{25} ( 23) (2008) 235009, {\tt arXiv:0807.0854}.
\bibitem{DA2}
W. Javed, J. Abbas,   A. Övgün,  \textit{Deflection angle of photon from magnetized black hole and effect
of nonlinear electrodynamics}, Eur. Phys. J. C, \textbf{79}  (2019) 694, {\tt arXiv:1908.09632}.
\bibitem{N}
G. W. Gibbons, M. Vyska, \textit{The application of Weierstrass elliptic functions to
Schwarzschild null geodesics}, Class. Quantum Grav. \textbf{29}  (2012) 065016, {\tt arXiv:1110.6508}.
\bibitem{DATD}
A. Belhaj, H. Belmahi, M. Benali, A. Segui, \textit{Thermodynamics of AdS black holes from deflection angle formalism}, Phys. Lett. B  \textbf{817} (2021) 136313.
\bibitem{DA6}
T. Ono, A. Ishihara,  H. Asada, \textit{Gravitomagnetic bending angle of light with finite-distance corrections in stationary axisymmetric spacetimes}, Phys. Rev. D \textbf{96} (2017) 104037, {\tt  arXiv:1704.05615}.



\bibitem{DA5}
R. Kumar, S. G. Ghosh, A. Wang, \textit{Shadow cast and deflection of light by charged rotating regular black holes}, Phys. Rev. D \textbf{100} (2019) 124024, {\tt arXiv:1912.05154}.

\bibitem{DA4}
A. Ishihara, Y. Suzuki, T. Ono, T. Kitamura,  H. Asada, \textit{Gravitational bending angle of light for finite distance and the Gauss-Bonnet theorem}, Phys. Rev. D \textbf{94} (2016) 084015, {\tt arXiv:1604.08308}
\bibitem{DA3}
B. E. Panah, K. Jafarzade, S. H. Hendi, \textit{ Charged 4D Einstein-Gauss-Bonnet-AdS Black Holes:
Shadow, Energy Emission, Deflection Angle and Heat Engine}, Nucl. Phys. B \textbf{961} (2020) 115269, {\tt arXiv:2004.04058}.
\bibitem{DAR1}
M. M. Caldarelli, G. Cognola, D. Klemm, \textit{Thermodynamics of Kerr-Newman-AdS Black Holes and Conformal Field Theories}, Class. Quant. Grav. \textbf{26} (2009) 195011, {\tt arXiv:hep-th/9908022}.

\bibitem{O}
P. Sharma, H. Nandan, R. Gannouji, R. Uniyal, A. Abebe, \textit{Deflection of light by a rotating black hole
surrounded by “quintessence”},  Int. J. Mod. Phys. A \textbf{35} (2020) 2050155, {\tt arXiv:1911.00372}.

\bibitem{54}
R. Adam, \textit{Frequency-dependent effects of gravitational lensing within plasma}, Monthly
Notices of the Royal Astronomical Society, \textbf{451} (2015) 4536, {\tt arXiv:1505.06790}.
\bibitem{56}
A. Abdujabbarov, B. Toshmatov, Z. Stuchlik, B. Ahmedov, \textit{ Shadow of the rotating black
hole with quintessential energy in the presence of plasma}, Int. J. Mod. Phys. D \textbf{26} (2016) 1750051,  {\tt arXiv:1512.05206}.



















































\end{thebibliography}
 \end{document}